\documentclass[conference]{IEEEtran}
\IEEEoverridecommandlockouts
\usepackage{cite}
\usepackage{amsmath,amssymb,amsfonts}
\usepackage{algorithmic}
\usepackage{graphicx}
\usepackage{textcomp}
\usepackage{xcolor}
\usepackage{multirow}
\def\BibTeX{{\rm B\kern-.05em{\sc i\kern-.025em b}\kern-.08em
    T\kern-.1667em\lower.7ex\hbox{E}\kern-.125emX}}

\usepackage{float}
\usepackage{hyperref}
\usepackage{gensymb}

\def\bfp{\mathbf p}

\begin{document}

\title{Deep Sequence-to-Sequence Models for\\ GNSS Spoofing Detection\\
\thanks{The work was partially supported by the project TM04000019 "Robust
Navigation System" of the Technology Agency of the Czech Republic.}
}

\author{\IEEEauthorblockN{1\textsuperscript{st} Jan Zelinka}
\IEEEauthorblockA{\textit{Department of Cybernetics} \\
\textit{University of West Bohemia}\\
Pilsen, Czech Republic \\
zelinka@kky.zcu.cz}
\and
\IEEEauthorblockN{2\textsuperscript{nd} Oliver Kost}
\IEEEauthorblockA{\textit{Department of Cybernetics} \\
\textit{University of West Bohemia}\\
Pilsen, Czech Republic \\
kost@kky.zcu.cz}
\and
\IEEEauthorblockN{3\textsuperscript{rd} Marek Hr\'uz}
\IEEEauthorblockA{\textit{Department of Cybernetics} \\
\textit{University of West Bohemia}\\
Pilsen, Czech Republic \\
mhruz@kky.zcu.cz}
}

\maketitle

\begin{abstract}
We present a data generation framework designed to simulate spoofing attacks and randomly place attack scenarios worldwide. We apply deep neural network-based models for spoofing detection, utilizing Long Short-Term Memory networks and Transformer-inspired architectures. These models are specifically designed for online detection and are trained using the generated dataset. Our results demonstrate that deep learning models can accurately distinguish spoofed signals from genuine ones, achieving high detection performance. The best results are achieved by Transformer-inspired architectures with early fusion of the inputs resulting in an error rate of 0.16\%.
\end{abstract}


\section{Introduction}

Unencrypted civilian global navigation satellite system (GNSS) signals are vulnerable to spoofing attacks, which pose a significant threat. Detecting such attacks is crucial for ensuring the security. However, obtaining real-world spoofing attack data is challenging, and acquiring a diverse dataset with heterogeneous scenarios from different locations across the globe is even more difficult.

To address this issue, we designed and implemented a data generation framework capable of simulating two distinct types of spoofing attacks. Our generator places attack scenarios at random locations on Earth and accounts for various signal conditions, including the presence of missing satellite signals. This enables the creation of a diverse and realistic dataset suitable for training machine learning models.

We applied deep neural network-based models to detect spoofing attacks, utilizing both Long Short-Term Memory (LSTM) networks and architectures inspired by modern Transformers. We have specifically designed the models to function as online spoofing detectors. Using our generated dataset, we trained and evaluated the models, assessing their effectiveness in distinguishing clean signals from spoofed ones. Our results demonstrate the feasibility of applying deep learning techniques to high-quality online spoofing detection.

\section{Related Works}
Various machine learning and deep learning approaches have been proposed for detecting spoofing attacks. Several studies have explored the use of neural networks, demonstrating their effectiveness \cite{9656691, 10.1145/3427228.3427254, NAYFEH2023103085, 9845684, 10.1007/978-3-030-59016-1_28}.
Among these, multi-layer perceptron models have been employed \cite{9634951, 10.1007/978-3-030-95467-3_38, Almadhor2025}. LSTM networks have been widely utilized for anomaly detection \cite{9217708, rs14194925}. Some works combine LSTM networks with convolutional neural networks (CNN) \cite{JIANG2022791} or using CNN-based approaches~\cite{s22239412, 10.1007/978-3-030-95467-3_38}.

\section{Nominal/Unspoofed PSR Signals}
The GNSS measurement of corrected\footnote{Corrections of satellite clock error, relativistic effect, ionosphere, and troposphere delays, etc., based on user-defined-model are applied.} pseudo-range (PSR) for the $l$-th satellite at time instant $k$ under nominal (i.e., unspoofed) conditions has the following form \cite{Gr:13}:
\begin{align}
	\rho^l_k &= || \bfp^l_k - \bfp_k || + b_k + \xi^l_k + \sigma^l_k, \label{PSR}
\end{align}
for $l=1, 2, \cdots, L_k$, where $\bfp_k$ is the position and $ b_k $ is the clock bias of the receiver, $\bfp^l_k$ is the position of the $l$-th satellite, and $L_k$ is the number of satellites visible from the receiver position at time instant $k$.

The PSR contains two types of noise, $\sigma^l_k$ and $\xi^l_k$, where both have zero-mean and are independent between satellites.
However, the noise $\sigma^l_k$ is independent in time (white) and the noise $ \xi^l_k $ is time-correlated. The time correlation is attributed to gradual changes in the atmosphere through which the GNSS signals pass.
The variances $\sigma^l_k$ and $\xi^l_k$ are functions of the elevation angle of the $l$-th satellite \cite{Gr:13,KoDuStDa:23}.

\section{Spoofing and Proposed Detection Concept}
\begin{figure}[H]
    \includegraphics[width=0.49\linewidth]{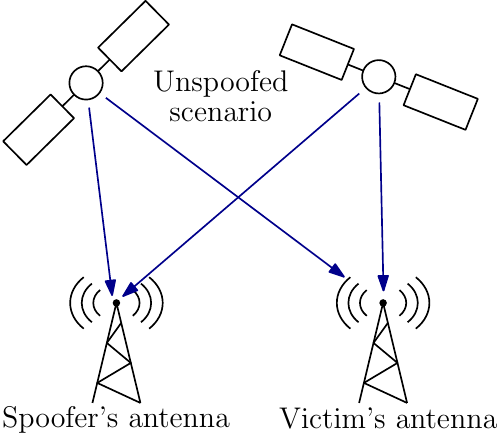}     
    \includegraphics[width=0.49\linewidth]{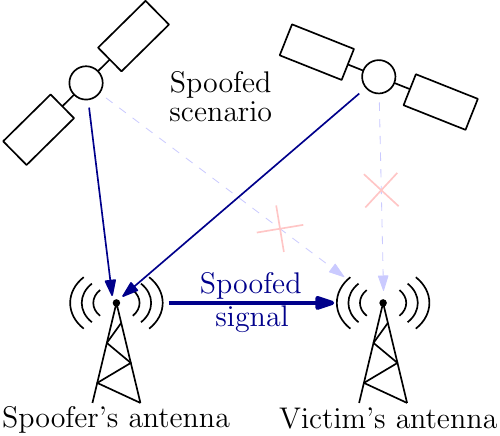}.
    \vspace{-5mm}
    \caption{Illustration of (un)spoofed GNSS signals.}
    \label{fig:spoofUnspoofIlust}
\end{figure}

A GNSS receiver processes signals transmitted by GNSS satellites orbiting at around 20,000 km high altitudes.
Therefore, the received signals are weak, which makes them susceptible to interference. The intentional interference~\footnote{For example, in 2011, Iran successfully managed to deceive a US drone with spoofed GNSS signals and let the drone land on its territory (\url{https://threat.technology/top-10-gps-spoofing-events-in-history/}). Currently, Ukraine, the Baltic countries, and Poland are witnessing massive jamming and spoofing of GNSS signals due to the military conflict triggered by Russia.} 
includes: 
\begin{itemize}
    \item \textit{Jamming}, where a strong signal overwhelms or blocks GNSS signals, prevents a victim's receiver from determining its position estimate. 
    This causes the navigation system to switch to another mode (e.g., inertial or terrain navigation \cite{Gr:13}).
    \item \textit{Spoofing}, where fake GNSS signals are sent to trick a victim's receiver into calculating an incorrect position, velocity, or time estimate (PVT solution). Spoofing can thus impair GNSS integrity, a key property determining the reliability, accuracy, and credibility of the provided estimates. This can result in fatal consequences, so it should be detected.
    Figure \ref{fig:spoofUnspoofIlust} illustrates GNSS signal spoofing.
\end{itemize}

\section{Generator of Spoofed PSR Signals: Targeted and Regional}
Consider two types of spoofing attacks: the first is \textit{targeted} at a single receiver, while the second affects all receivers in a given \textit{region} near the spoofer.
Both types of spoofing attacks modify GNSS signals received by an aircraft moving along a trajectory that follows the pattern described in \cite{ED259}. However, the position of the aircraft's trajectory and the runway is not fixed, and the aircraft lands from any direction at any location on Earth for different simulations. Moreover, consider that the aircraft follows a nominal/unspoofed trajectory during the entire flight, and the spoofing attack modifies only the GNSS signals received by the aircraft. Note that the spoofer knows his (static) position in both types of spoofing attacks.

\subsection{Targeted spoofing}
\begin{figure}[H]
    \includegraphics[width=1\linewidth]{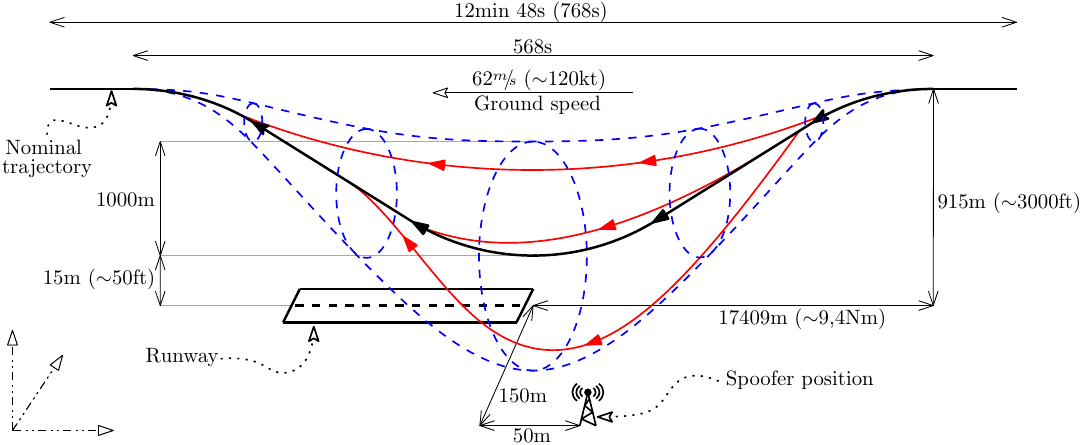}
    \vspace{-5mm}
    \caption{Illustration of targeted spoofing trajectories. The red lines are examples of spoofing trajectories, and the blue dashed lines indicate the area where spoofed trajectories might occur.}
    \label{pristani_vse_X}
\end{figure}

The targeted spoofing attack involves the spoofer attacking one specific victim and shifting his position via spoofed PSR signals. 
The shift of the spoofed trajectory from the nominal (unspoofed) trajectory is smooth.
The duration of the spoofed attack ranges from $100s$ to $568s$. In the middle of the nominal trajectory, the spoofed trajectory is shifted from the nominal one by $300m$ to $1000m$ as is illustrated in Figure \ref{pristani_vse_X}.
To carry out a targeted spoofing attack, the spoofer needs to estimate the victim's position (e.g., using radar) as well as the GNSS time and PSR correlated noises (e.g., by measuring nominal GNSS signals and using a Kalman filter on the spoofer side), and potentially other estimates. 
Since the victim's PVT solution may not exhibit significant deviations before and during the spoofing attack, the errors in these estimates contribute to distinguishing the spoofed GNSS signal from the nominal signal.

\subsection{Regional spoofing}
\begin{figure}[H]
    \includegraphics[width=1\linewidth]{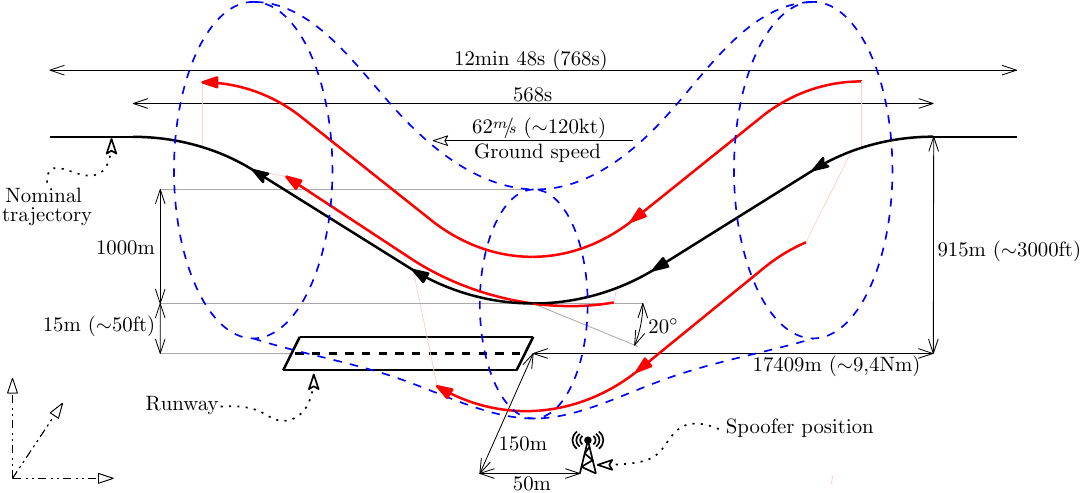}
    \vspace{-5mm}
    \caption{Illustration of regional spoofing trajectories. The red lines are examples of spoofing trajectories, and the blue dashed lines indicate the area where spoofed trajectories might occur.}
    \label{pristani_vse_Y}
\end{figure}

The regional spoofing attack does not target a single victim but the entire region where the spoofer transmits one set of (untargeted) spoofed PSR signals for all victims occurring in this region. 
The spoofed trajectory is generated by taking a part of the nominal (unspoofed) trajectory of duration $100s$ to $568s$.
The whole part of the trajectory is shifted from the middle of the nominal trajectory by $300m$ to $1000m$. 
In addition, the spoofed trajectory is horizontally rotated by $\pm 20^\circ$ around the middle of the nominal trajectory as is illustrated in Figure \ref{pristani_vse_Y}.
Compared to a targeted spoofing attack, a regional one needs almost\footnote{For example, as in this article, GNSS time estimation can be used.} no estimates. Therefore, the spoofed GNSS signal may be more difficult to detect during the attack, but on the other hand, it can be easily detected at the beginning and end of the spoofing attack, where step changes occur (see Figure \ref{pristani_vse_Y}).

\section{Input Signals Processing}

In this section, we describe the preprocessing of input signals, enabling the use of neural network-based models. In our context, $\rho$ represents the PSR measurement, which is the apparent distance between a satellite and a receiver based on the time delay of the received signal~\eqref{PSR}.

There are several key challenges in using $\rho$ as an input for an artificial neural network: signal is not always present, its magnitude varies significantly, and the most relevant information is encoded in its small fluctuations over time despite its overall large magnitude. These issues prevent $\rho$ from being used directly as an input.

We addressed the issue of missing signals by not only using a single value of $\rho$ but also incorporating an indicator variable with a value of 0 or 1 to denote its presence or absence.

Using the raw $\rho$ value as input would directly lead to gradient vanishing or exploding due to significant differences in magnitude. Standard normalization techniques, such as batch normalization, do not fully resolve this issue, as they reduce magnitude variation but also diminish valuable information contained in small temporal variations in the network's signals. Employing the second difference mitigates this problem by emphasizing subtle fluctuations while removing the influence of the overall large magnitude of $\rho$.

\begin{figure}[H]
    \includegraphics[width=\linewidth]{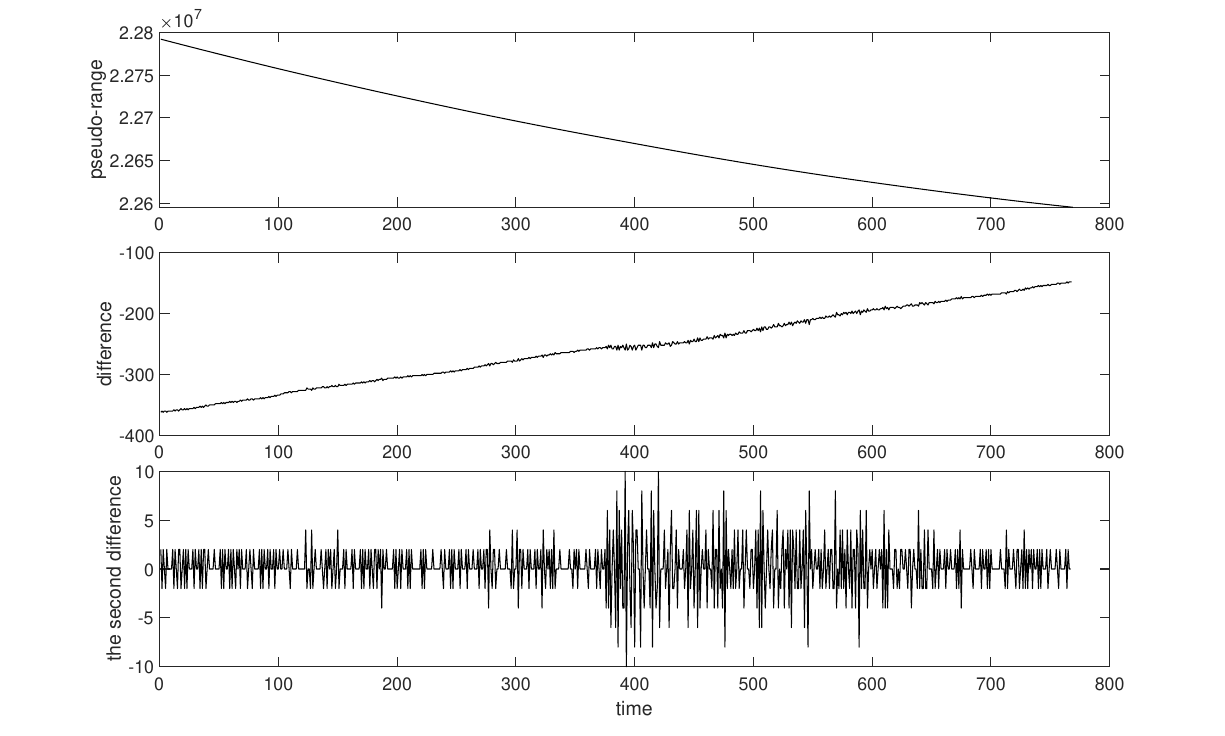}
    \caption{An example illustrating PSR and its differences.}
    \label{fig:rhodrhoddrho_example}
\end{figure}

In the context of a vehicle using GNSS, the second difference of $\rho$ is correlated with the acceleration of the vehicle.
This means that analyzing the second difference of $\rho$ allows us to detect variations in the vehicle’s motion, such as false sudden acceleration that could be caused by a spoofing attack (see Figure \ref{fig:rhodrhoddrho_example}).

The second difference retains a significant amount of useful information but also amplifies noise, including quantization noise. We do not apply explicit smoothing, assuming that employed machine learning models can inherently learn and perform the necessary smoothing during training. The second difference can still have a large magnitude (usually caused by some irregularity of the signal), thus we apply a transformation to compress its range $f(x) = \mathrm{sign}(x) \cdot \mathrm{ln}(1 + |x|)$.
For small values of $x$, it behaves approximately linearly, ensuring smooth transitions near zero. For large values, it grows logarithmically, reducing the impact of extreme magnitudes while preserving the sign of $x$.

Modern models of sequence processing neural networks require the inputs to be embedded into a vector representation. Instead of ineffective projecting a single value $y = f(x)$ through a linear transformation to construct an embedding vector, we quantize it with $N$ values and express it as a probability distribution over a set of learnable quantized values $q_i$. These probabilities $p(y)$ are computed as follows:
\begin{equation}
p(y) = \mathrm{softmax}([\lambda_1\|y - q_1\|, \dots, \lambda_N\|y - q_N\|]),
\end{equation}
where $\lambda_i$ are learnable scaling parameters that control the sharpness of the distribution.
The key advantage of this approach is that the quantization process can be pretrained as an autoencoder once and subsequently used across different models. 
We performed this pretraining using a linear layer for reconstruction and the mean squared error criterion. The pretrained quantizer was then employed in all our experiments.

\section{Approaches to Signal Fusion}

We have data from a variable number of satellites. The data can be processed independently for each satellite, with the results subsequently combined, or they can be aggregated into a single vector of constant dimension by applying padding to match the maximum number of satellites, allowing for processing using a single model. The first approach is referred to as late fusion, while the second as early fusion.

In the case of early fusion, indicators of signal presence are included in the inputs, allowing the applied models to determine which parts of each signal to use. For late fusion with transformers, these indicators can be incorporated in two ways: either by including them in the input or by using them for attention masking to directly prevent the model from processing missing signals.

The advantage of late fusion is its ability to process a variable number of satellites, while early fusion enables a direct combination of signals from different satellites during processing.

\section{Sequence-to-Sequence Model}

We compared two methods for capturing long-range dependencies in sequential data: LSTM networks and the Multi-Head Attention (MHA) mechanism. We use only the self-attention component of the MHA mechanism as described in the literature~\cite{NIPS2017_3f5ee243}.

While LSTMs inherently maintain causality by processing information sequentially from past to future, the MHA mechanism enforces causality through masking. This masking not only ensures that each time step attends only to previous positions but also enables the exclusion of missing signals from computation.
This allows input signals to be processed without the need to explicitly include missing signal indicators in the inputs.

The MHA mechanism enables the processing of data from multiple satellites even in late fusion by extending the attention mechanism to attend not only to different time steps of the same satellite but also to data from other satellites. To reduce computational demands, we restrict the attention mechanism for different satellites to only consider data from the same time step, excluding interactions between different satellites across different time steps. However, in models with multiple layers utilizing MHA, this restriction does not strictly prevent the processing of cross-satellite temporal dependencies, as information can still be propagated through successive layers.

When computing attention, the primary objective of using positional embeddings is to utilize not only variations in the embedded signal values at different positions but also the positional differences themselves. Therefore, in our case, it is essential to employ two distinct positional embeddings: one for time and another for the satellite index.

In the case of early fusion with MHA, the embedding vectors incorporate data from all satellites. This ensures that interactions between different satellites are already encoded within the input representation, eliminating the need for explicit cross-satellite attention during processing.

\begin{figure}
    \centering
    \includegraphics[trim={0cm 1cm 0 1.4cm}, width=0.75\linewidth]{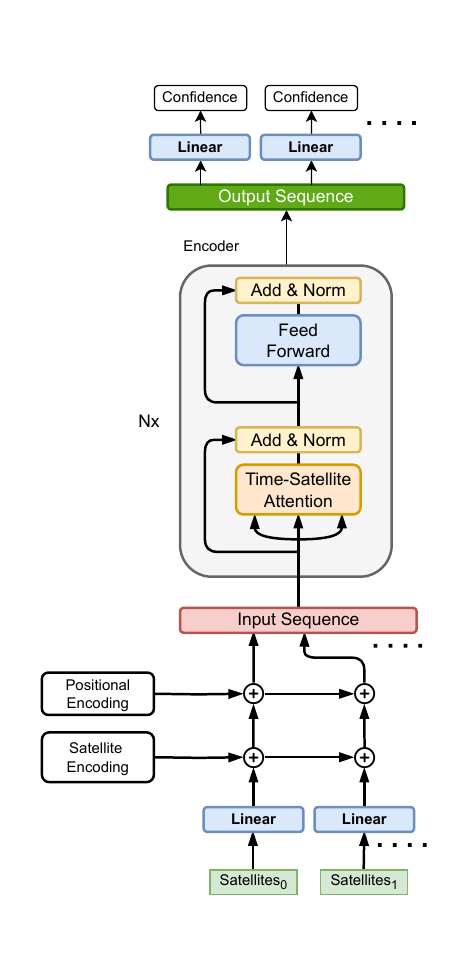}
    \caption{Illustration of the proposed multi-headed attention-based module for satellite sequence data processing. Input is a sequence of satellite embeddings in every time step. Output is the confidence of spoofing in that time step.}
    \label{fig:module_arch}
\end{figure}

We utilized a standard module (see Fig.~\ref{fig:module_arch}) in which the input is first processed by a selected layer, either an LSTM or a MHA mechanism. The processed output is then combined with the original input through a residual connection and subsequently normalized using layer normalization. This is followed by a feed-forward network, which is also enhanced with a residual connection and layer normalization, applied in the same manner.
This approach leverages residual connections and layer normalization to mitigate vanishing gradient issues.
By incorporating either LSTM or MHA, the model can effectively capture  sequential dependencies.

The detection output is a continuous score ranging from 0 to 1 for each time point.
Instead of one score value, our models generate two values, apply a softmax function, and train using cross-entropy loss.
This approach offers the advantage of explicitly modeling the competition between classes, which can improve the robustness of the training process.
The second score is disregarded after applying the softmax function, as the result is fully determined by the first one.

In the case of early fusion, generating a resultant score poses no issue, as a vector with a fixed number of elements can be directly converted into a score. However, in the case of late fusion, the number of vectors varies. To address this, we employ a simplified attention mechanism.
This mechanism computes an individual output and an associated weight for each vector separately. The final output is computed as follows: first, the individual weights $w_i$ are collected, and the softmax function is applied. These weights are then used to compute a weighted average of the individual outputs $y_i$. Finally, the softmax function is applied to generate the resultant scores $y$. Formaly for $n$ satellites:
$v = \mathrm{softmax}([w_1, \dots w_n])$, $y = \mathrm{softmax}([v_1 y_1, \dots, v_n y_n])$.
This approach ensures that the contributions of different inputs are adaptively weighted.

\section{Experiments and Results}

We generated a substantial corpus of over 67,000 sequences of PSR for training and an additional 1,000 sequences for testing. The test set contains 21.96 \% of spoofed signal.
We believe that this dataset will provide the necessary foundation for training models, which often require extensive data.

Our dataset provides a unique opportunity for a specialized training approach that cannot be replicated using datasets derived solely from real-world measurements. During each gradient computation, the model receives two sets of $\rho$ measurements: one containing a spoofing attack within a restricted time interval and another with identical measurements but without any attack. This setup enables the training process to develop models that are more explicitly focused on distinguishing relevant information.

\begin{figure}[H]
    \includegraphics[width=0.9\linewidth]{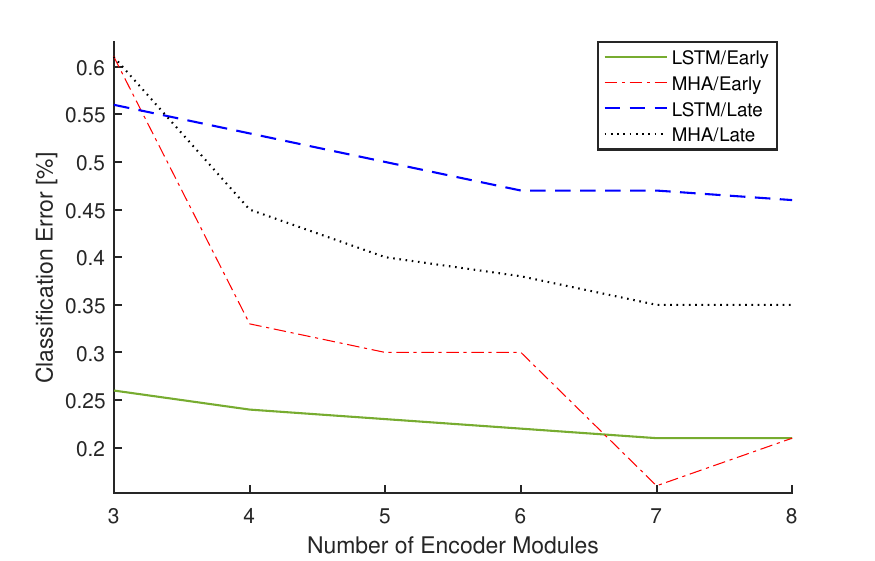}     
    \caption{Classification error for different model configurations and number of encoder modules ($\mathbf{N}$ in Fig.~\ref{fig:module_arch}).}
    \label{fig:errvslayers}
\end{figure}

In our setup, we used an embedding dimension of 128. The number of neurons in the hidden layers of the feed-forward networks was 1024, and we utilized 8 heads in the MHA mechanism.  Our experiments consisted of four configurations: LSTM-based models with early fusion, MHA-based models with early fusion, LSTM-based models with late fusion, and MHA-based models with late fusion. For each configuration, we optimized the number of layers required for effective detection. Specifically, we evaluated encoder architectures with module counts (see Figure~\ref{fig:module_arch}) ranging from 1 to 8 (see Figure~\ref{fig:errvslayers}).
A summary of our results are in Table~\ref{tab:results}.
Performance was evaluated by calculating the classification error.

\begin{table}[!t]
\renewcommand{\arraystretch}{1.1}
\caption{Summary of the results comparing the two proposed model architectures and the two fusion strategies.}
\label{tab:results}
\centering
\begin{tabular}{|c|c|c|c|}
\hline
\bfseries Model & \bfseries Targeted att. & \bfseries  Regional att. & \bfseries Total\\
\hline\hline

\begin{tabular}{c}
LSTM\\
Early F.
\end{tabular}
&
\begin{tabular}{c}
err = 0.44 \% \\
fa = 0.25 \% \\
md = 0.19 \% \\
\end{tabular}
&
\begin{tabular}{c}
\textbf{err = 0.01 \%} \\
fa = 0.01 \% \\
\textbf{md = 0.00 \%} \\
\end{tabular}
&
\begin{tabular}{c}
err = 0.21 \% \\
fa = 0.12 \% \\
md = 0.09 \% \\
\end{tabular}
\\
\hline

\begin{tabular}{c}
MHA\\
Early F.
\end{tabular}
&
\begin{tabular}{c}
\textbf{err = 0.31 \%} \\
\textbf{fa = 0.11 \%} \\
md = 0.20 \% \\
\end{tabular}
&
\begin{tabular}{c}
err = 0.03 \% \\
\textbf{fa = 0.00 \%} \\
md = 0.03 \% \\
\end{tabular}
&
\begin{tabular}{c}
\textbf{err = 0.16 \%} \\
\textbf{fa = 0.05 \%} \\
md = 0.11 \% \\
\end{tabular}
\\
\hline

\begin{tabular}{c}
LSTM\\
Late F.
\end{tabular}
&
\begin{tabular}{c}
err = 0.70 \% \\
fa = 0.50 \% \\
md = 0.20 \% \\
\end{tabular}
&
\begin{tabular}{c}
err = 0.26 \% \\
fa = 0.26 \% \\
\textbf{md = 0.00 \%} \\
\end{tabular}
&
\begin{tabular}{c}
err = 0.47 \% \\
fa = 0.38 \% \\
md = 0.10 \% \\
\end{tabular}
\\
\hline

\begin{tabular}{c}
MHA\\
Late F.
\end{tabular}
&
\begin{tabular}{c}
err = 0.46 \% \\
fa = 0.34 \% \\
\textbf{md = 0.11 \%} \\
\end{tabular}
&
\begin{tabular}{c}
err = 0.26 \% \\
fa = 0.26 \% \\
\textbf{md = 0.00} \% \\
\end{tabular}
&
\begin{tabular}{c}
err = 0.35 \% \\
fa = 0.30 \% \\
\textbf{md = 0.05 \%} \\
\end{tabular}
\\
\hline

\end{tabular}
\end{table}

In our experiments, we set a threshold of 0.5 for the predicted scores to identify spoofing attacks. We evaluated the model's performance by calculating the classification error, false alarm (FA) rate, and missed detection (MD) rate.
Since our dataset includes an equal distribution of two distinct types of spoofing attacks, we assessed the model on three different sets: a subset containing only targeted attacks, a subset consisting only of regional attacks, and the entire dataset. Targeted attacks alter signals slowly and in small increments, while regional attacks modify signals in a single step.

Our results indicate that MHA-based models outperform LSTM-based models. Additionally, the early fusion strategy proved to be the most effective. It is expected that regional attacks are detected with a higher error rate.

\section{Conclusion}

In this paper, we presented a generator specifically designed for spoofing attacks. This generator was used to create a large dataset containing a diverse range of examples of various spoofing attacks. We utilized this dataset to train deep machine learning models for spoofing detection, addressing challenges such as missing signals. We developed a specialized input signal coding system and a modified attention mechanism. Our experiments demonstrated that spoofing attacks could be detected by online detectors with high accuracy. However, it needs to be noted that the demonstration was performed on simulated data and the real-world application should be investigated further.

\bibliographystyle{IEEEtran}
\bibliography{literatura} 

\end{document}